\documentclass[aps,twocolumn]{revtex4-1}
\usepackage{amsmath}
\usepackage{graphicx}

\begin{document}

\title{Non-Hermitian dynamics of Cooper pair splitter }
\author{E. S. Ma}
\author{Z. Song}
\email{songtc@nankai.edu.cn}
\affiliation{School of Physics, Nankai University, Tianjin 300071, China}
\begin{abstract}
We propose a non-Hermitian model for Cooper pair splitters, in which the
process of electron tunneling into electrodes is characterized by
non-Hermitian terms. We find that across a broad range of parameters, the
energy levels consistently remain real, and coalescing states are always
present. The Coulomb repulsion between electrons in a quantum dot affects the
order of the coalescing states. This gives rise to two distinct dynamic
behaviors: (i) when the initial state is an empty state, the final state
supports a nonzero electron-escaping rate; (ii) the electron-escaping rate
is zero for a single-electron initial state. In the former case, our exact
solutions reveal that the average electron-escaping rate vanishes along a
set of hyperbolic curves in the plane of the chemical potentials of the two
quantum dots. The stability of the results in the presence of disordered perturbation is also investigated. Our findings pave the way for investigating  Cooper pair splitters within the framework of non-Hermitian quantum mechanics.
\end{abstract}

\maketitle

\section{Introduction}

Rigorous results for a model Hamiltonian play an important role in physics
and sometimes open new avenues for exploration in the field. The exact
solution to the quantum harmonic oscillator has played a crucial role in the
history of physics. It stands as a key concept in traditional quantum
mechanics and continues to be a cornerstone for modern research and
applications within the field. One recent example is the discovery of the
solution for the non-Hermitian harmonic oscillator, which serves as the
foundation for {$\mathcal{PT}$}-symmetric quantum mechanics {\cite%
{Bender,Bender1,Bender2,Bender3}}. In traditional quantum mechanics, the
Hamiltonian being Hermitian is a fundamental postulate that ensures that the
energy spectrum is real and that the system undergoes unitary time evolution 
{\cite{DAMQ}. }In contrast, the real spectrum and unitary dynamics are not
required to be tied together in an open system. In recent years,
non-Hermitian physics {\cite{Ali1,Ali2,Ali3,Ali4,Ali5}} has attracted much
attention from various research areas \cite{longhi2009bloch, jin2010physics,
lee2014heralded, jin2017topological, qin2020discrete, ashida2020non,
jin2021symmetry, zhang2021observation, rohn2023classical,
liang2023observation}. Non-Hermitian quantum physics is particularly
relevant for describing open quantum systems that interact with their
environment, leading to phenomena such as damping probability and the
appearance of nonorthogonal states.

Cooper pair splitters are devices that can separate Cooper pairs \cite%
{lesovik2001electronic,recher2001}, which are pairs of electrons that are
bound together in a superconducting state, into individual electrons while
maintaining their quantum entanglement. It has been explored in various
experimental setups, including those based on nanowires \cite%
{hofstetter2009cooper,hofstetter2011finite,das2012high,fulop2014local,fulop2015magnetic,baba2018cooper,scherubl2022from,kurtossy2022parallel,bordoloi2022spin,wang2022singlet,jong2023single}%
, carbon nanotubes \cite%
{herrmann2010carbon,schindele2012near,herrmann2012arXiv,bruhat2018circuit},
graphene \cite%
{tan2015cooper,borzenets2016high,tan2021thermoelectric,pandey2021ballistic},
and semiconductor quantum dots \cite{jong2023single,bordin2023arXiv}. These
devices have shown promising results in generating entangled electrons,
which are essential for quantum applications. Theoretical investigations are
usually performed within the framework of Hermitian quantum mechanics \cite%
{hiltscher2011adiabatic,brange2021dynamic,eldridge2010superconducting,
walldorf2020noise,brange2014adiabatic}. However, it is a typical open system 
\cite{sauret2004quantum} in which the superconductor and electrodes can be
considered as the electron source and drains. From a theoretical point of
view, a non-Hermitian Hamiltonian is a suitable candidate to characterize
this system. Traditionally, Cooper pair splitters have been modeled using
Hermitian quantum mechanics. However, it is crucial to propose a paradigm
shift by incorporating non-Hermitian effects into our analysis. These
effects are essential for capturing the nonreciprocal nature of the
electron tunneling process.

In this work, we introduce a non-Hermitian model to analyze the behavior of
a Cooper pair splitter, a device critical for the manipulation of electron
pairs in open quantum systems. By incorporating non-Hermitian terms into the
Hamiltonian, we aim to provide a more comprehensive understanding of the
electron tunneling process and its implications for the energy levels and
dynamics of the system. Through the exact solutions of the non-Hermitian
Hamiltonian, we find that the energy levels in our non-Hermitian model
consistently remain real, indicating the stability of the system. Moreover,
we identify the presence of coalescing states, which are pivotal in
understanding the dynamics of the electron-pair splitting process. Such
dynamics are exclusive to non-Hermitian systems. We also find that the
Coulomb repulsion between electrons in a quantum dot is a significant factor
affecting the order of coalescing states. Our exact results demonstrate that
the average electron-escaping rate decreases along a set of hyperbolic
curves in the chemical potential plane of the two quantum dots. This finding
offers a clear indication of the interplay among non-Hermitian effects,
Coulomb repulsion, and dynamics.

This paper is organized as follows. In Section \ref{Model and phase diagram}%
, we introduce the Hamiltonian of the non-Hermitian Cooper pair splitter and
provide the complete exact solutions. In Section \ref{Two types of dynamics}%
, we discuss two types of dynamics for two representative initial states.
Two observables are defined to characterize the dynamics. The robustness of
the dynamics under the time-dependent disorder perturbation is investigated
in Section \ref{Stability in the presence of disordered perturbation}.
Finally, we draw conclusions in Section \ref{Summary}. Some details of the
derivations are provided in the Appendix.

\begin{figure*}[tbh]
\centering
\includegraphics[width=0.8\textwidth]{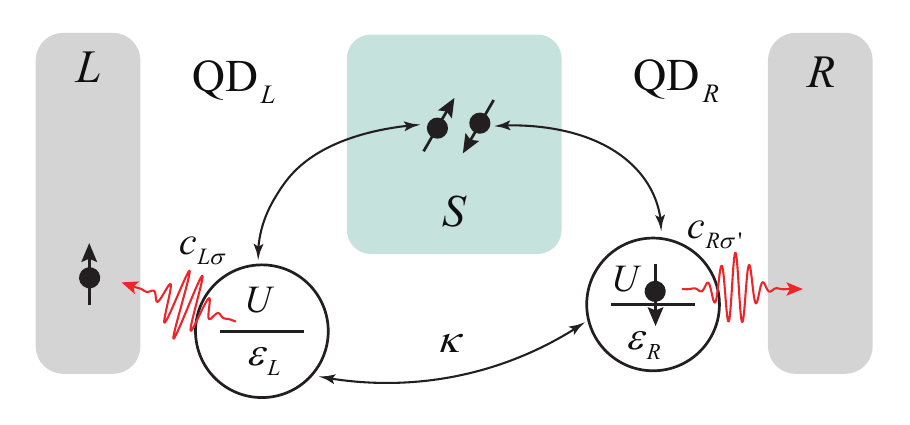}
\caption{The schematic depicts a Cooper pair splitter as described by the
Hamiltonian in Eq. (\protect\ref{H}). $S$ represents a superconductor, which
serves as the source of Cooper pairs. Each quantum dot is represented by
black circles, with respective chemical potentials $\protect\varepsilon _{L}$
and $\protect\varepsilon _{R}$. The Hubbard on-site interaction is denoted
by $U$. The hopping strength between two quantum dots is denoted by $%
\protect\kappa $. $L$ and $R$ represent the left and right electrodes,
respectively. The process of electrons escaping from quantum dots is
characterized by the non-Hermitian term in Eq. (\protect\ref{nonH}).}
\label{fig1}
\end{figure*}

\begin{figure*}[tbh]
\centering
\includegraphics[width=0.9\textwidth]{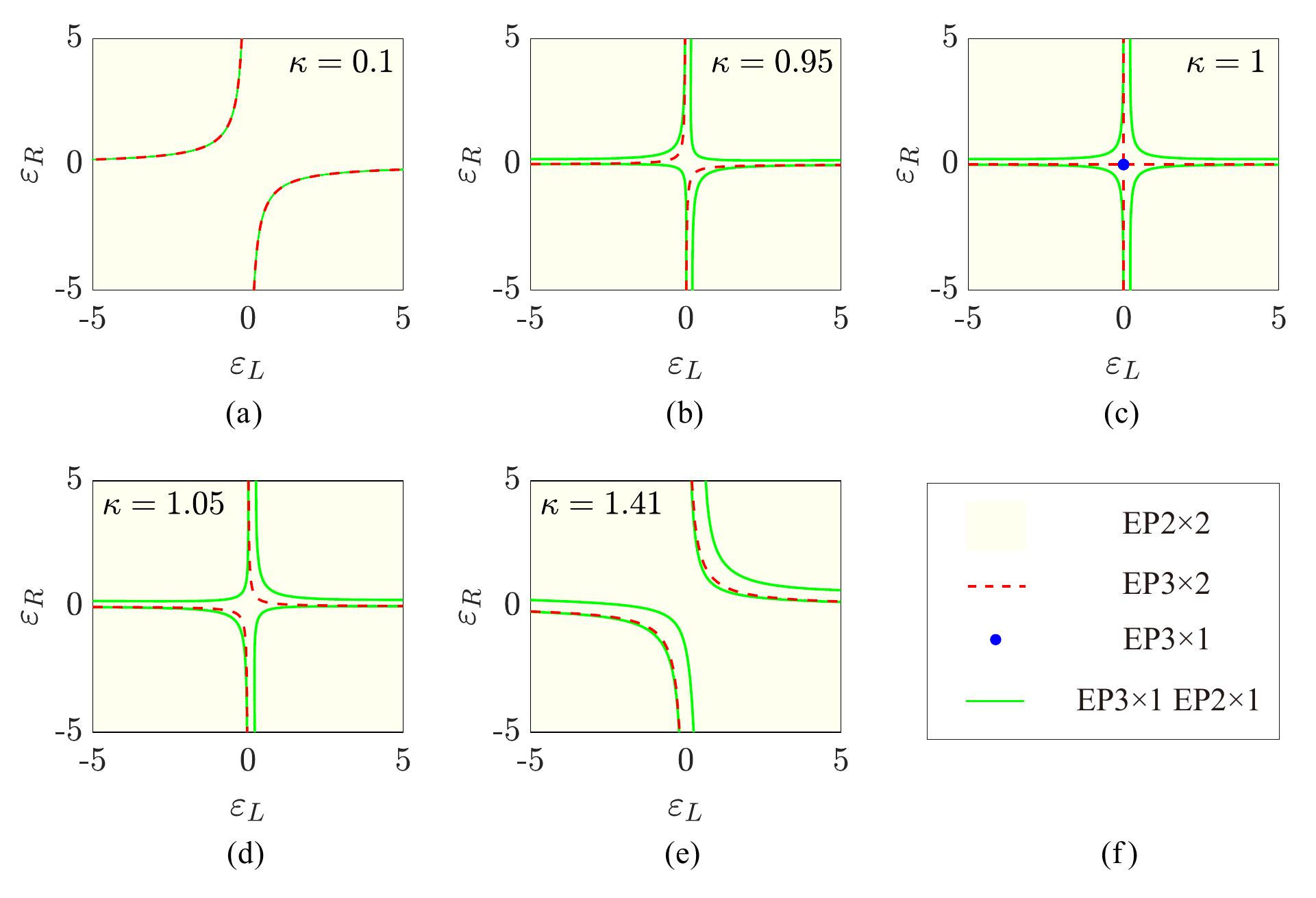}
\caption{The phase diagram for the Hamiltonian represented by Eq. (\protect
\ref{H}) is depicted on the $\protect\varepsilon _{L}-\protect\varepsilon %
_{R}$ parameter plane for several representative values of $\protect\kappa $
and $U$ ($\protect\gamma =1$). The configurations corresponding to the
coalescing energy levels are indicated by color-coded regions or curves.
For $U=20$, the green solid lines indicate a $2$nd-order EP and a $3$%
rd-order EP. For $U=\infty $, the red dashed lines indicate two $3$rd-order
EPs.}
\label{fig2}
\end{figure*}

\section{Model and phase diagram}

\label{Model and phase diagram}

We begin with a non-Hermitian Cooper pair splitter, and the Hamiltonian is 
\begin{equation}
H=H_{1}+H_{2},  \label{H}
\end{equation}%
with 
\begin{eqnarray}
H_{1} &=&-\kappa \sum\limits_{\sigma }c_{L\sigma }^{\dag }c_{R\sigma
}-\gamma d_{S}^{\dag }+\mathrm{h.c.}  \notag \\
&&+\sum\limits_{\ell \sigma }\varepsilon _{\ell }c_{\ell \sigma }^{\dag
}c_{\ell \sigma }+U\sum\limits_{\ell }n_{\ell \uparrow }n_{\ell \downarrow },
\end{eqnarray}%
and 
\begin{equation}
H_{2}=\lambda \sum\limits_{\ell \sigma }c_{\ell \sigma },  \label{nonH}
\end{equation}%
where $\varepsilon _{\ell }$ represents the energy level of quantum dots,
which can be tuned by external gates and $\ell =L,R$. $\sigma =\uparrow
,\downarrow $ is the index of the electron spin. $d_{S}^{\dag }=\left(
c_{L\downarrow }^{\dag }c_{R\uparrow }^{\dag }-c_{L\uparrow }^{\dag
}c_{R\downarrow }^{\dag }\right) /\sqrt{2}$ describes a singlet state. The
amplitudes for Cooper pair splitting and elastic cotunneling are dominated
by $\gamma $ and $\kappa $, respectively. $U$ is the interdot Coulomb
potential and we consider $U\gg \kappa ,\gamma ,\lambda $ for simplicity. $%
H_{2}$ is the non-Hermitian term, which is responsible for transmitting
split electrons from quantum dots to electrodes. A schematic of the
system is displayed in Fig. \ref{fig1}.

The correlation between two electrons stems from the on-site Hubbard
interaction $U$, which is crucial for understanding phenomena such as the
Mott transition, antiferromagnetism, and superconductivity. To
characterize the strong electron correlations, an effective Hamiltonian, such
as the $t-J$ Hamiltonian, can be derived by using the Schrieffer-Wolff
transformation \cite{bravyi2011schrieffer}, with the transformation
generator depending on $t/U$ and excluding the possibility for electrons to
doubly occupy a single site \cite{eckle2019models}.

The effective Hamiltonian reads%
\begin{eqnarray}
H_{\mathrm{eff}} &=&\sum\limits_{\ell \sigma }\varepsilon _{\ell }\tilde{c}%
_{\ell \sigma }^{\dag }\tilde{c}_{\ell \sigma }-\kappa \sum\limits_{\sigma
}\left( \tilde{c}_{L\sigma }^{\dag }\tilde{c}_{R\sigma }+\mathrm{h.c.}%
\right) +\lambda \sum\limits_{\ell \sigma }c_{\ell \sigma }  \notag \\
&&+J_{\mathrm{eff}}\left( \mathbf{s}_{L}\cdot \mathbf{s}_{R}-\frac{1}{4}%
\right)-\gamma \left( d_{S}^{\dag }+d_{S}\right), \label{Heff}
\end{eqnarray}%
where the exchange coupling strength $J_{\mathrm{eff}}=4U\kappa ^{2}/\left[
U^{2}-\left( \varepsilon _{L}-\varepsilon _{R}\right) ^{2}\right] $, and $%
s_{\ell }^{\alpha }$\ ($\alpha =x,y,z$) is the spin-$1/2$ operator. The
operators $\tilde{c}_{\ell \sigma }^{\dag }$ and $\tilde{c}_{\ell ^{\prime
}\sigma ^{\prime }}$\ satisfy the relations 
\begin{equation}
\left\{ \tilde{c}_{\ell \sigma }^{\dag },\tilde{c}_{\ell ^{\prime }\sigma
^{\prime }}\right\} =\delta _{\ell \ell ^{\prime }}\delta _{\sigma \sigma
^{\prime }},\tilde{c}_{\ell \sigma }^{\dag }\tilde{c}_{\ell \sigma ^{\prime
}}^{\dag }=0.
\end{equation}%
Based on the basis set 
\begin{equation}
\left( 1,d_{S}^{\dag },c_{L\uparrow }^{\dag },c_{R\uparrow }^{\dag
},c_{L\downarrow }^{\dag },c_{R\downarrow }^{\dag }\right) \left\vert
0\right\rangle ,
\end{equation}%
where $\left\vert 0\right\rangle $ is the vacuum state and satisfies $%
c_{\ell \sigma }\left\vert 0\right\rangle =0$, the matrix representation $h$
of $H_{\mathrm{eff}}$ and the solution of equation%
\begin{equation}
h\psi _{n}=\epsilon _{n}\psi _{n},h^{\dag }\phi _{n}=\epsilon _{n}^{\ast
}\phi _{n},
\end{equation}%
($n\in \lbrack 1,6]$) are obtained in the appendix A. The
analysis indicates that coalescing states \cite{moiseyev2011non} exist, and
the number of these states depends on the system parameters. Specifically,
finite values of $U$ and infinite values of $U$ lead to different degrees of
energy level degeneracy and result in distinct EPs. The details are
presented in the appendix, and the structure of the solutions is
demonstrated as a phase diagram in Fig. \ref{fig2}. Notably, except for the
parameters at several hyperbolic curves in the phase diagram, the coalescing
state lacks the components of the vacuum state $\left\vert 0\right\rangle $\
and singlet pair state $d_{S}^{\dag }\left\vert 0\right\rangle $, while
other eigenstates are superposition of all the basis states. This is
crucial for the dynamics that will be discussed in the following section.

\begin{figure*}[tbh]
\centering
\includegraphics[width=0.9\textwidth]{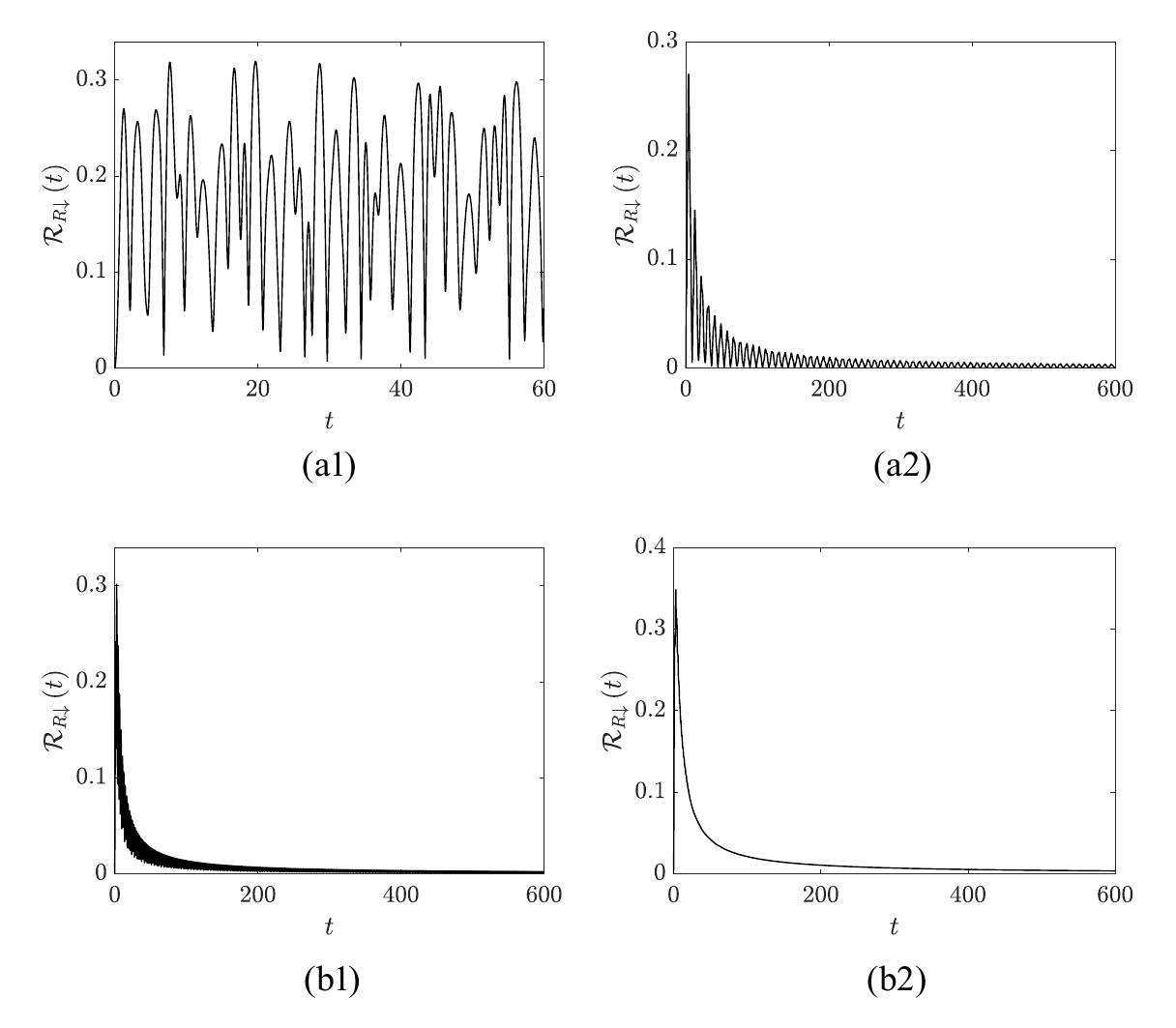}
\caption{The numerical results of the electron-escaping rate defined in Eq. (%
\protect\ref{R}). For $U=20$, the initial state for (a1) and (a2) is $%
\left\vert 0\right\rangle $ and $c_{L\uparrow }^{\dag }\left\vert
0\right\rangle $, respectively. As expected, (a1) exhibits quasi-Hermitian
dynamics and (b1) exhibits EP dynamics. Other parameters are $\protect\gamma %
=1$, $\protect\lambda =1$, $\protect\kappa =1$, $\protect\varepsilon _{L}=1$%
, and $\protect\varepsilon _{R}=2$. The plots in (b1) and (b2) are the
results for the same initial state $\left\vert 0\right\rangle $ but with $%
U\rightarrow \infty $ for (b1) and $U=20$ for (b2). The parameters for (b1)
are $\protect\gamma =1$, $\protect\lambda =1$, $\protect\kappa =1.41$, $%
\protect\varepsilon _{L}=0.5$, and $\protect\varepsilon _{R}=1.98$,
corresponding to the point located at the red dashed line in Fig. \protect
\ref{fig2}(e). The parameters for (b2) are $\protect\gamma =1$, $\protect%
\lambda =1$, $\protect\kappa =1.41$, $\protect\varepsilon _{L}=3.689$, and $%
\protect\varepsilon _{R}=0.239$, corresponding to the point located at the
green solid line in Fig. \protect\ref{fig2}(e). The results agree with our
predictions, corresponding to 2nd-order and 3rd-order EP dynamics,
respectively.}
\label{fig3}
\end{figure*}

\begin{figure*}[tbh]
\centering
\includegraphics[width=0.9\textwidth]{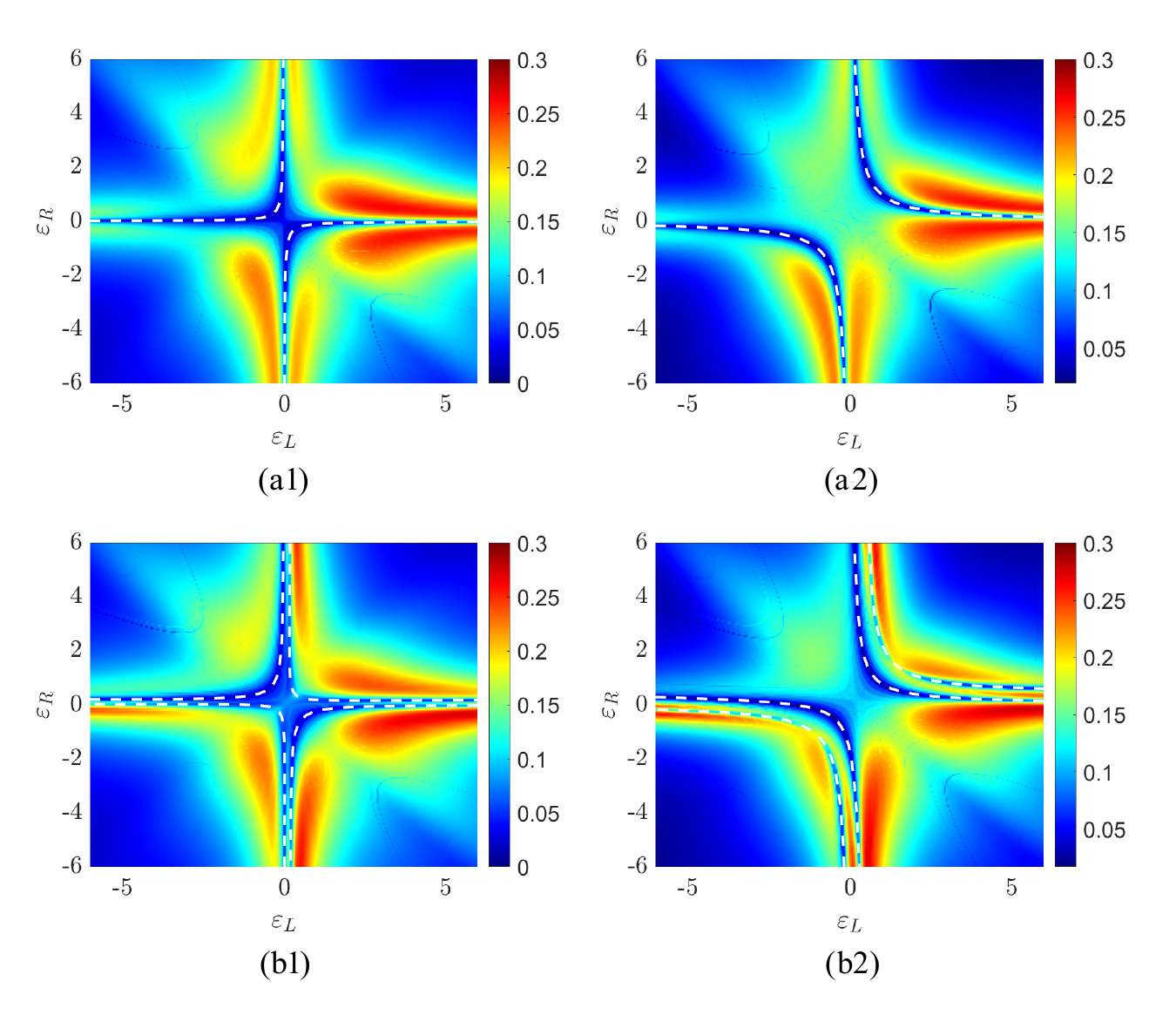}
\caption{Color contour plots of the numerical results for the average
current $\mathcal{I}_{R\downarrow }$ defined in Eq. (\protect\ref{I1}). We
take $U\rightarrow \infty $\ for (a1) and (a2) and $U=20$ for (b1) and (b2), 
$\protect\kappa =0.95$ for (a1) and (b1) and $\protect\kappa =1.41$ for (a2)
and (b2). The initial state is $\left\vert 0\right\rangle $ and $\protect\gamma =1$, $\protect\lambda =1$ for all the cases. The white dashed lines represent the EP lines obtained from the exact results. We can see that the current is exactly zero along the EP lines for infinite $U$, and the current is relatively smaller along the EP lines for finite $U$. These results indicate that the current can demonstrate EP lines both for finite and infinite $U$. }
\label{fig4}
\end{figure*}

\begin{figure*}[tbh]
\centering
\includegraphics[width=0.9\textwidth]{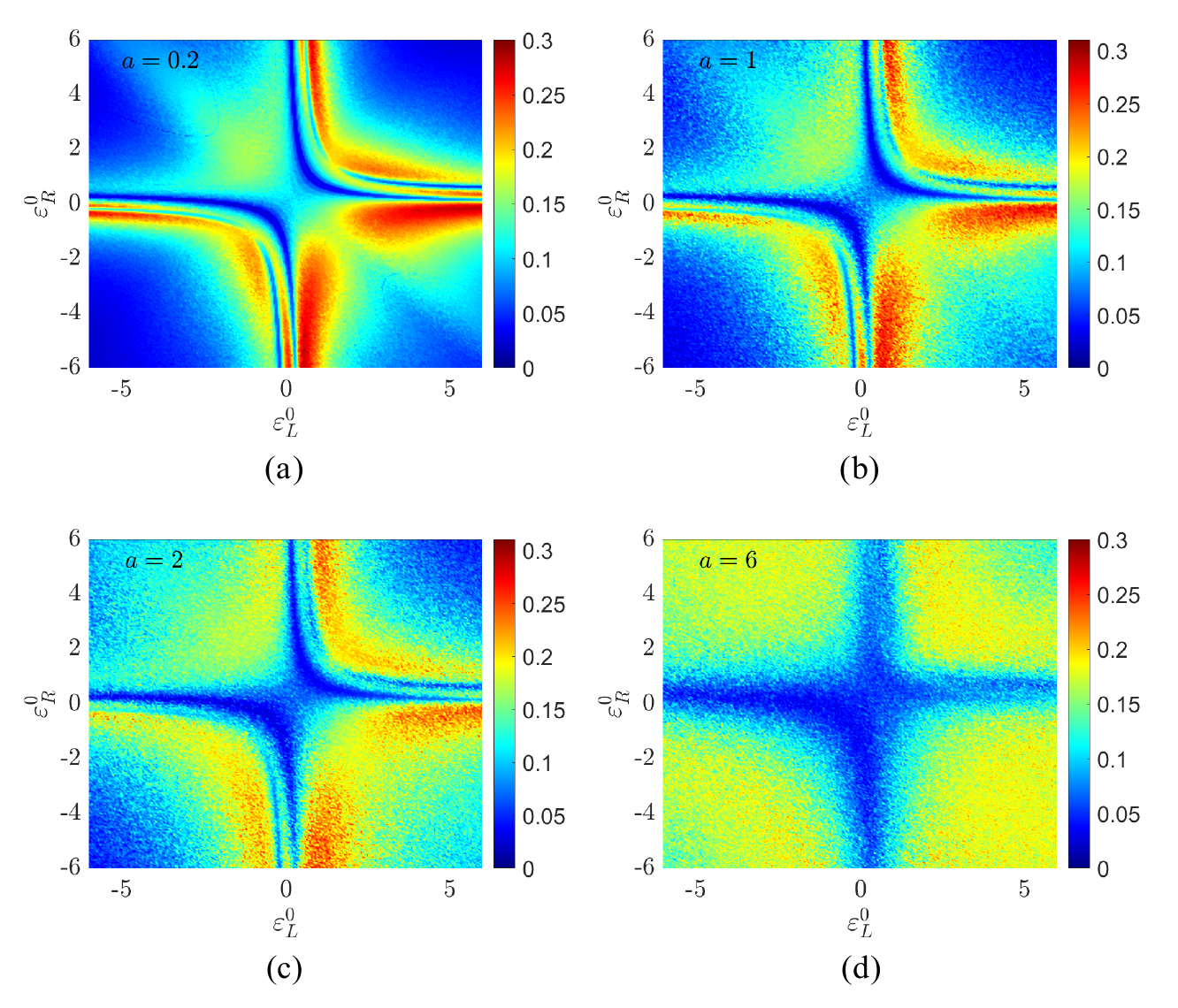}
\caption{Color contour plots of the average current for the time evolution
of the initial state $\left\vert 0\right\rangle $ are driven by the
Hamiltonian with time-dependent chemical potentials in Eq. (\protect\ref%
{disorder}). The numerical results are obtained from Eq. (\protect\ref{I2})
with representative values of $a$. The other parameters are the same as those in
Fig. \protect\ref{fig4}(b2). This indicates that the exact results are robust
against a certain range of disorder perturbations, and the difference between the cases with infinite and finite $U$ disappears as $a$\ increases.}
\label{fig5}
\end{figure*}

\section{Two types of dynamics}

\label{Two types of dynamics}

In this section, we will discuss the dynamics of such a non-Hermitian
system. In general, there are two types of dynamics of a non-Hermitian system with a full real spectrum: (i) EP dynamics, in which the initial state is driven by a
Jordan block. The corresponding final state approaches the coalescing state
after long-term evolution. (ii) Quasi-Hermitian dynamics, the initial state
is the superposition of a set of biorthonormal eigenstates and 
coalescing states. The evolved state is oscillatory, but the Dirac
probability is nonconservative.

The solution of the present non-Hermitian Cooper-pair-splitter system
indicates that two types of dynamics are involved (see Appendix B) within
the yellow region in the phase diagram. In the following, we introduce the
electron-escaping rate, denoted by the following formula: 
\begin{equation}
\mathcal{R}_{\ell \sigma }\left( t\right) =\left\vert \left\langle \psi
\left( t\right) \right\vert c_{\ell \sigma }\left\vert \psi \left( t\right)
\right\rangle \right\vert ,  \label{R}
\end{equation}%
for an evolved state. Additionally, we define the average current as: 
\begin{equation}
\mathcal{I}_{\ell \sigma }=\frac{1}{T}\int_{0}^{T}\mathcal{R}_{\ell \sigma
}\left( t\right) dt,  \label{I1}
\end{equation}%
to characterize the efficiency of the Cooper pair splitter. Here we consider
the time evolution of two specific initial states for the yellow region in
the phase diagram.

(i) The initial state is the vacuum state $\left\vert \psi \left( 0\right)
\right\rangle =\left\vert 0\right\rangle $. Within the yellow region, where $%
\xi _{1,2}\neq 0$, we have%
\begin{equation}
\left\vert \psi \left( 0\right) \right\rangle =\sum_{i=1}^{4}\alpha
_{i}\left\vert \psi _{i}\right\rangle ,
\end{equation}%
where the eigenstate $\left\vert \psi _{i}\right\rangle $ denotes%
\begin{equation}
\psi _{i}^{\text{T}}\left( 1,d_{S}^{\dag },c_{L\uparrow }^{\dag
},c_{R\uparrow }^{\dag },c_{L\downarrow }^{\dag },c_{R\downarrow }^{\dag
}\right) ^{\text{T}}\left\vert 0\right\rangle .
\end{equation}%
The initial state is the superposition of two biorthonormal eigenstates $%
\left\{ \left\vert \psi _{1}\right\rangle ,\left\vert \psi _{2}\right\rangle
\right\} $\ and two coalescing states $\left\{ \left\vert \psi
_{3}\right\rangle ,\left\vert \psi _{4}\right\rangle \right\} $. Then the
evolved state is%
\begin{equation}
\left\vert \psi \left( t\right) \right\rangle =e^{-iHt}\left\vert \psi
\left( 0\right) \right\rangle =\sum_{i=1}^{4}\alpha _{i}e^{-i\epsilon
_{i}t}\left\vert \psi _{i}\right\rangle ,
\end{equation}%
which is periodic, with its period determined by the energy levels $\left\{
\epsilon _{i}\right\} $. Accordingly, we have%
\begin{equation}
\mathcal{R}\left( t\right) \propto \left\vert \xi _{1}f_{1}\left( t\right)
+\xi _{2}f_{2}\left( t\right) \right\vert ,
\end{equation}%
where $f_{1}(t)$ and $f_{2}(t)$ are periodic functions.

(ii) The initial state is a single electron occupied state, $\left\vert \psi
\left( 0\right) \right\rangle =c_{\ell \sigma }^{\dag }\left\vert
0\right\rangle $. Within the yellow region, where $\xi _{1,2}\neq 0$, we
have 
\begin{equation}
\left\vert \psi \left( t\rightarrow \infty \right) \right\rangle \propto
\left( \mu \left\vert \psi _{3}\right\rangle +\nu \left\vert \psi
_{4}\right\rangle \right) t.
\end{equation}%
Accordingly, we have%
\begin{equation}
\mathcal{R}\left( t\rightarrow \infty \right) \rightarrow 0,
\end{equation}%
due to the fact%
\begin{equation}
\left\langle \psi _{i}\right\vert c_{\ell \sigma }\left\vert \psi
_{j}\right\rangle =0,
\end{equation}%
if $i,j=3,4.$ Now, we turn to the case beyond the yellow region, where the
parameters satisfy Eq. (\ref{finite}) for $J_{\mathrm{eff}}\neq 0$ and
Eq. (\ref{infinite}) for $J_{\mathrm{eff}}=0$. In these situations, the time
evolution of $\left\vert \psi \left( 0\right) \right\rangle =\left\vert
0\right\rangle $ involves the Jordan block, which obeys EP dynamics. Accordingly,
we have%
\begin{equation}
\mathcal{R}\left( t\rightarrow \infty \right) \rightarrow 0,
\end{equation}%
which is distinct from the results observed in the yellow region.

Numerical simulations of the electron-escaping rate $\mathcal{R}_{\ell
\sigma }\left( t\right) $ for the system, using typical parameters, are
performed to demonstrate our results. The evolved states, corresponding to
two distinct types of initial states, can be obtained through numerically
exact diagonalization. The results are presented in Fig. \ref{fig3}, which
accords with our predictions.

Obviously, the average current is always zero for the initial state $c_{\ell
\sigma }^{\dag }\left\vert 0\right\rangle $ for any given system parameters.
In contrast, it can be nonzero for the initial state $\left\vert
0\right\rangle $ and nearly vanishes when a $2$nd-order EP becomes a $3$rd-order
EP according to curves in Eq. (\ref{finite}) for $J_{\mathrm{eff}}\neq 0$ and completely vanishes in Eq. (%
\ref{infinite}) for $J_{\mathrm{eff}}=0$. Fig. \ref{fig4} shows the
numerical results of the average current in the $\varepsilon _{L}$-$\varepsilon
_{R}$ plane. We can see that the observable $\mathcal{I}_{\ell \sigma }$\
clearly demonstrates that the higher-order EPs appear in the non-Hermitian
Cooper-pair-splitter system. Importantly, it also indicates that $\mathcal{I}%
_{\ell \sigma }$\ reaches its maximum in the vicinity of the zeros of $%
\varepsilon _{L}$ or $\varepsilon _{R}$.

\section{Stability in the presence of disordered perturbation}

\label{Stability in the presence of disordered perturbation}

In this section, we focus on the situation involving a specific type of
disorder and hope that the dynamics are robust.

We consider the case where there are time-dependent random perturbations in
the energy levels of quantum dots, which appear in the form 
\begin{equation}
\varepsilon _{R,L}\left( t\right) =\varepsilon _{R,L}^{0}+\mathrm{ran}\left(
-a,a\right) _{t}.  \label{disorder}
\end{equation}%
The notation $\mathrm{ran}\left( -a,a\right) _{t}\ $represents a uniformly
distributed random variable in the interval $\left( -a,a\right) $, resulting
in the parameter $\varepsilon _{R,L}\left( t\right) $\ at time $t$. Due to
the presence of the perturbation, it is difficult to obtain the exact
evolved results. Numerical simulation is an efficient method for investigating
this problem, and computations are performed on a uniform time mesh for
discretization, i.e., stroboscopic time evolution. For a given initial state,
the time-evolved state is computed by using 
\begin{equation}
\left\vert \psi \left( t_{n}\right) \right\rangle =\mathcal{T}%
\prod\limits_{i=1}^{n}e^{-iH\left( t_{i-1}\right) \left(
t_{i}-t_{i-1}\right) }\left\vert \psi \left( 0\right) \right\rangle ,
\end{equation}%
where $\mathcal{T}$ is the time-order operator and $t_{i}-t_{i-1}=\delta t$
is a constant. Accordingly, the energy levels take the form 
\begin{equation}
\varepsilon _{R,L}\left( t_{l}\right) =\varepsilon _{R,L}^{0}+\mathrm{ran}%
\left( -a,a\right) _{l},
\end{equation}%
and the average current is computed by using%
\begin{equation}
\mathcal{I}_{\ell \sigma }\approx \frac{1}{T}\sum_{i=1}^{M}\left\vert
\left\langle \psi \left( t_{i}\right) \right\vert c_{\ell \sigma }\left\vert
\psi \left( t_{i}\right) \right\rangle \right\vert \delta t.  \label{I2}
\end{equation}

The numerical results for the cases with finite Hubbard $U$ and a range of
representative values of $a$ are plotted in Fig. \ref{fig5}. These results
indicate that they remain robust against a certain range of perturbations. As $a$
increases, the effect of finite $U$ diminishes.

\section{Summary}

\label{Summary}

In summary, we introduced a non-Hermitian model to study the behavior of a
Cooper pair splitter. The model incorporates non-Hermitian terms to describe
the electron tunneling process into electrodes, capturing the nonreciprocal
nature of this phenomenon. Our results reveal that within a wide parameter
space, the energy levels of the system remain real and exhibit coalescing
states. The presence of Coulomb repulsion between electrons in a quantum dot
influences the ordering of these coalescing states, leading to two distinct
dynamic behaviors. When the system starts in an empty state, the final
state allows for a nonzero electron-escaping rate, indicating that
electrons can be transferred to the electrodes. Conversely, if the system
begins with a single-electron state, the electron-escaping rate is zero,
suggesting a trapped state with no electron transfer. The exact solutions
demonstrate that the average electron-escaping rate decreases along
hyperbolic curves in the chemical potential plane of the two quantum dots,
providing a unique signature of non-Hermitian dynamics. The findings not
only contribute to the understanding of the Cooper pair splitter within the
framework of non-Hermitian quantum mechanics but also open up new avenues
for further research. Future work will explore the practical implications of
these theoretical insights, including potential applications in quantum
computing and quantum information processing.

\acknowledgments This work was supported by the National Natural Science
Foundation of China (under Grant No. 12374461).

\section*{Appendix}
\label{Appendix}

In this appendix, we present a derivation for the solution to the
Hamiltonian in Eq. (\ref{H}), based on which the phase diagram in Fig. \ref%
{fig2} can be obtained. In addition, we also provide a brief introduction to
the dynamics driven by a Jordan block, which forms the foundational theory
for our investigation in Sec. \ref{Two types of dynamics}.

\subsection*{A. The solution to the Hamiltonian}

\label{A}  
\setcounter{equation}{0} \renewcommand{\theequation}{A\arabic{equation}}

In this subsection, we derive the effective Hamiltonian in Eq. (\ref{Heff}) and
provide its solutions. In the limit of a large Coulomb potential, where $%
U\gg \lambda ,\kappa ,\gamma $, the double-occupied states $c_{L\uparrow
}^{\dag }c_{L\downarrow }^{\dag }\left\vert 0\right\rangle $ and $%
c_{R\uparrow }^{\dag }c_{R\downarrow }^{\dag }\left\vert 0\right\rangle $\
can be neglected when considering low-energy eigenstates. To obtain
the effective Hamiltonian for low-energy eigenstates, we consider the
subspace constructed by the following basis%
\begin{equation}
\left( 1,d_{S}^{\dag },c_{L\uparrow }^{\dag },c_{R\uparrow }^{\dag
},c_{L\downarrow }^{\dag },c_{R\downarrow }^{\dag }\right) \left\vert
0\right\rangle ,  \label{basis}
\end{equation}%
which forms a set of eigenstates of the sub-Hamiltonian $H_{0}$,%
\begin{equation}
H_{0}=\sum\limits_{\ell \sigma }\varepsilon _{\ell }c_{\ell \sigma }^{\dag
}c_{\ell \sigma }+U\sum\limits_{\ell }n_{\ell \uparrow }n_{\ell \downarrow }.
\end{equation}%
Taking the hopping term%
\begin{equation}
H^{\prime }=-\kappa \sum\limits_{\sigma }\left( c_{L\sigma }^{\dag
}c_{R\sigma }+\mathrm{h.c.}\right) ,
\end{equation}%
as a perturbation, the effective Hamiltonian of $H_{0}+H^{\prime }$\ can be
obtained as%
\begin{eqnarray}
&&H_{0}+H^{\prime }\rightarrow \sum\limits_{\ell \sigma }\varepsilon _{\ell }%
\tilde{c}_{\ell \sigma }^{\dag }\tilde{c}_{\ell \sigma }-\kappa
\sum\limits_{\sigma }\left( \tilde{c}_{L\sigma }^{\dag }\tilde{c}_{R\sigma }+%
\mathrm{h.c.}\right)  \notag \\
&&+J_{\mathrm{eff}}\left( \mathbf{s}_{L}\cdot \mathbf{s}_{R}-\frac{1}{4}%
\right) ,
\end{eqnarray}%
where the exchange coupling strength 
\begin{equation}
J_{\mathrm{eff}}=\frac{4U\kappa ^{2}}{U^{2}-\left( \varepsilon
_{L}-\varepsilon _{R}\right) ^{2}},
\end{equation}%
and the spin-$1/2$ operator is defined by%
\begin{equation}
s_{\ell }^{\alpha }=\frac{1}{2}\left( \tilde{c}_{\ell \uparrow }^{\dag },%
\tilde{c}_{\ell \downarrow }^{\dag }\right) \sigma ^{\alpha }\left( 
\begin{array}{c}
\tilde{c}_{\ell \uparrow } \\ 
\tilde{c}_{\ell \downarrow }%
\end{array}%
\right) ,
\end{equation}%
with $\sigma ^{\alpha }$ ($\alpha =x,y,z$) being the Pauli matrix. The operators 
$\tilde{c}_{\ell \sigma }^{\dag }$ and $\tilde{c}_{\ell ^{\prime }\sigma
^{\prime }}$\ satisfy the relations 
\begin{equation}
\left\{ \tilde{c}_{\ell \sigma }^{\dag },\tilde{c}_{\ell ^{\prime }\sigma
^{\prime }}\right\} =\delta _{\ell \ell ^{\prime }}\delta _{\sigma \sigma
^{\prime }},\tilde{c}_{\ell \sigma }^{\dag }\tilde{c}_{\ell \sigma ^{\prime
}}^{\dag }=0.
\end{equation}%
Together with the non-Hermitian and pairing\ terms we obtain the effective
Hamiltonian%
\begin{eqnarray}
H_{\mathrm{eff}} &=&\sum\limits_{\ell \sigma }\varepsilon _{\ell }\tilde{c}%
_{\ell \sigma }^{\dag }\tilde{c}_{\ell \sigma }-\kappa \sum\limits_{\sigma
}\left( \tilde{c}_{L\sigma }^{\dag }\tilde{c}_{R\sigma }+\mathrm{h.c.}%
\right)+\lambda \sum\limits_{\ell \sigma }c_{\ell \sigma }  \notag \\
&&+J_{\mathrm{eff}}\left( \mathbf{s}_{L}\cdot \mathbf{s}_{R}-\frac{1}{4}%
\right)-\gamma \left( d_{S}^{\dag }+d_{S}\right).
\end{eqnarray}

Now, we turn to the solution of the effective Hamiltonian, $H_{\mathrm{eff}}$%
. Based on the basis set in Eq. (\ref{basis}), the matrix representation of $%
H_{\mathrm{eff}}$ is obtained as

\begin{equation}
h=\left( 
\begin{array}{cccccc}
0 & -\gamma & \lambda & \lambda & \lambda & \lambda \\ 
-\gamma & \varepsilon _{R}+\varepsilon _{L}-J_{\mathrm{eff}} & 0 & 0 & 0 & 0
\\ 
0 & \lambda /\sqrt{2} & \varepsilon _{L} & -\kappa & 0 & 0 \\ 
0 & \lambda /\sqrt{2} & -\kappa & \varepsilon _{R} & 0 & 0 \\ 
0 & -\lambda /\sqrt{2} & 0 & 0 & \varepsilon _{L} & -\kappa \\ 
0 & -\lambda /\sqrt{2} & 0 & 0 & -\kappa & \varepsilon _{R}%
\end{array}%
\right) .
\end{equation}%
The solution to the system satisfies the Schrodinger equations 
\begin{equation}
h\psi _{n}=\epsilon _{n}\psi _{n},h^{\dag }\phi _{n}=\epsilon _{n}^{\ast
}\phi _{n},
\end{equation}%
($n\in \lbrack 1,6]$). It is easy to check that the explicit forms of the
eigenvectors and eigenvalues are

\begin{equation}
\psi _{1,2}=\left( 
\begin{array}{c}
-\sqrt{2}\gamma \xi _{1,2}/\epsilon _{1,2} \\ 
\sqrt{2}\xi _{1,2} \\ 
\lambda \left( \kappa +\varepsilon _{R}-\epsilon _{1,2}\right) \\ 
\lambda \left( \kappa +\varepsilon _{L}-\epsilon _{1,2}\right) \\ 
-\lambda \left( \kappa +\varepsilon _{R}-\epsilon _{1,2}\right) \\ 
-\lambda \left( \kappa +\varepsilon _{R}-\epsilon _{1,2}\right)%
\end{array}%
\right) ,
\end{equation}%
and%
\begin{eqnarray}
\psi _{3} &=&\psi _{5},\psi _{4}=\psi _{6},  \notag \\
\psi _{3,4} &=&\left( 
\begin{array}{c}
0 \\ 
0 \\ 
\kappa +\varepsilon _{R}-\epsilon _{3,4} \\ 
\kappa +\varepsilon _{L}-\epsilon _{3,4} \\ 
-\left( \kappa +\varepsilon _{R}-\epsilon _{3,4}\right) \\ 
-\left( \kappa +\varepsilon _{L}-\epsilon _{3,4}\right)%
\end{array}%
\right) ,
\end{eqnarray}%
with%
\begin{eqnarray}
\epsilon _{1,2} &=&\frac{\varepsilon _{L}+\varepsilon _{R}-J_{\mathrm{eff}%
}\pm \sqrt{\left( \varepsilon _{L}+\varepsilon _{R}-J_{\mathrm{eff}}\right)
^{2}+4\gamma ^{2}}}{2},  \notag \\
\epsilon _{3} &=&\epsilon _{5},\epsilon _{4}=\epsilon _{6},  \notag \\
\epsilon _{3,4} &=&\frac{\varepsilon _{L}+\varepsilon _{R}\pm \sqrt{\left(
\varepsilon _{L}-\varepsilon _{R}\right) ^{2}+4\kappa ^{2}}}{2},
\end{eqnarray}%
where the parameter is%
\begin{equation}
\xi _{1,2}=\epsilon _{1,2}J_{\mathrm{eff}}-\gamma ^{2}-\varepsilon
_{L}\varepsilon _{R}+\kappa ^{2}.
\end{equation}%
In parallel, we have 
\begin{equation}
\phi _{1,2}=\left( 
\begin{array}{c}
\xi _{1,2} \\ 
-\xi _{1,2}\epsilon _{1,2}/\gamma \\ 
\lambda \left( \varepsilon _{R}-\epsilon _{1,2}+\kappa \right) \\ 
\lambda \left( \varepsilon _{L}-\epsilon _{1,2}+\kappa \right) \\ 
\lambda \left( \varepsilon _{R}-\epsilon _{1,2}+\kappa \right) \\ 
\lambda \left( \varepsilon _{L}-\epsilon _{1,2}+\kappa \right)%
\end{array}%
\right) ,
\end{equation}%
and 
\begin{eqnarray}
\phi _{3} &=&\phi _{5},\phi _{4}=\phi _{6},  \notag \\
\phi _{3,4} &=&\left( 
\begin{array}{c}
0 \\ 
0 \\ 
\varepsilon _{R}-\epsilon _{3,4}+\kappa \\ 
\varepsilon _{L}-\epsilon _{3,4}+\kappa \\ 
\varepsilon _{R}-\epsilon _{3,4}+\kappa \\ 
\varepsilon _{L}-\epsilon _{3,4}+\kappa%
\end{array}%
\right) .
\end{eqnarray}

We note that all the energy levels $\left\{ \epsilon _{n}\right\} $\ are
real, and $\psi _{3}$ and $\psi _{4}$ are two $2$nd-order
coalescing states. This can also be confirmed by the identities%
\begin{equation}
\phi _{3}^{\dag }\psi _{3}=\phi _{4}^{\dag }\psi _{4}=0.
\end{equation}%
That is, the biorthogonal modes of two states are zero.

We conclude that, in general, the solution to the system consists of
two real levels and two $2$nd-order coalescing levels, except for
the following case. The equation of $\xi _{1}=0$ or $\xi _{2}=0$
corresponds to the curve in the parameter $\varepsilon _{L}$-$\varepsilon
_{R}$ plane, which is described by the following equation

\begin{equation}
J_{\mathrm{eff}}^{2}\left( \varepsilon _{L}\varepsilon _{R}-\kappa
^{2}\right)+J_{\mathrm{eff}}\left( \varepsilon _{L}+\varepsilon _{R}\right)
\zeta +\zeta ^{2}=0.  \label{finite}
\end{equation}%
Here, the coefficient is defined as $\zeta =\kappa ^{2}-\gamma
^{2}-\varepsilon _{L}\varepsilon _{R}$. Under this condition, we find that
one element of $\left\{ \epsilon _{1},\epsilon _{2}\right\} $ is equal to
one element of $\left\{ \epsilon _{3},\epsilon _{4}\right\} $, indicating the
coalescence between one state of $\left\{ \psi _{1},\psi _{2}\right\} $ and
one state of $\left\{ \psi _{3},\psi _{4}\right\} $. Therefore, the solution
of the system consists of one $2$nd-order coalescing level and one $3$%
rd-order coalescing level for $\xi_{1,2}=0 $ when $J_{\mathrm{eff}}$ is not
equal to zero.

When we consider the limit as $U$ approaches infinity, and $J_{\mathrm{eff}}$
approaches $0$, the energy levels remain real. The solution of the system
consists of two real levels and two $2$nd-order coalescing levels.
Particularly, at the curve 
\begin{equation}
\varepsilon _{L}\varepsilon _{R}=\kappa^{2}-\gamma ^{2},  \label{infinite}
\end{equation}
in the parameter $\varepsilon _{L}$-$\varepsilon _{R}$\ plane, we have $%
\zeta =0$, and then 
\begin{equation}
\psi _{1}=\psi _{3}=\psi _{5},\psi _{2}=\psi _{4}=\psi _{6},
\end{equation}%
and%
\begin{equation}
\phi _{1}=\phi _{3}=\phi _{5},\phi _{2}=\phi _{4}=\phi _{6}.
\end{equation}%
The solution of the system consists of two $3$rd-order coalescing levels,
which can be confirmed by the identities%
\begin{equation}
\phi _{1}^{\dag }\psi _{1}=\phi _{2}^{\dag }\psi _{2}=0.
\end{equation}

Accordingly, in the region where $\kappa ^{2}-\gamma ^{2}>0$, the hyperbolic
curves are in the first and third quadrants. Conversely, when $\kappa
^{2}-\gamma ^{2}<0$, they are in the second and fourth quadrants.
Specifically, under these conditions, two of the hyperbolic curves
degenerate to lines along two principal axes%
\begin{equation}
\varepsilon _{R}=0,\text{ or }\varepsilon _{L}=0,
\end{equation}%
at $\kappa =\pm \gamma $.

\subsection*{B. EP dynamics}

\label{B} \setcounter{equation}{0} \renewcommand{%
\theequation}{B\arabic{equation}}

In this subsection, we present a general formalism for the temporal
evolution of an initial state that includes a Jordan block. In non-Hermitian
systems, an EP is a parameter value at which two or more eigenvalues and
their corresponding eigenvectors coalesce. This is a point of non-Hermitian
degeneracy where the usual spectral decomposition of the operator fails
because the eigenvectors are no longer linearly independent.

Mathematically, an EP is always associated with a Jordan block corresponding
to a single eigenvalue $\lambda $. This block has the form:%
\begin{equation}
J_{d}=\left[ 
\begin{array}{ccccc}
\lambda & 1 & 0 & \cdots & 0 \\ 
0 & \lambda & 1 & \cdots & 0 \\ 
\vdots & \vdots & \ddots & \vdots & \vdots \\ 
0 & 0 & \cdots & \lambda & 1 \\ 
0 & 0 & \cdots & 0 & \lambda%
\end{array}%
\right] ,
\end{equation}%
which is a $d\times d$\ matrix, satisfying $\left( J_{d}-\lambda \right)
^{d}=0$.

Applying the above analysis to the solution for $h$, we have the following
results. For the case with two $2$nd-order EPs, there exist two auxiliary
states $\psi _{3,4}^{\mathrm{au}}$ that satisfy

\begin{equation}
\left( h-\epsilon _{3,4}\right) \psi _{3,4}^{\mathrm{au}}=\psi _{3,4}.
\end{equation}%
Then, for an initial state with nonzero components of $\psi _{3,4}^{\mathrm{au%
}}$, we have the long-term evolution 
\begin{equation}
\left\vert \psi \left( t\rightarrow \infty \right) \right\rangle \propto
\left( \mu \left\vert \psi _{3}\right\rangle +\nu \left\vert \psi
_{4}\right\rangle \right) t.
\end{equation}

On the other hand, for the case with two $3$rd-order EPs, there exists four
auxiliary states $\left\{ \psi _{j}^{\mathrm{au}},j\in \left[ 1,4\right]
\right\} $ obeying%
\begin{equation}
\left( h-\epsilon _{3}\right) \psi _{1}^{\mathrm{au}}=\psi _{3}^{\mathrm{au}%
},\left( h-\epsilon _{3}\right) \psi _{3}^{\mathrm{au}}=\psi _{3},
\end{equation}%
and 
\begin{equation}
\left( h-\epsilon _{4}\right) \psi _{2}^{\mathrm{au}}=\psi _{4}^{\mathrm{au}%
},\left( h-\epsilon _{4}\right) \psi _{4}^{\mathrm{au}}=\psi _{4}.
\end{equation}%
Then, for an initial state with nonzero components of $\psi _{1,2,3,4}^{%
\mathrm{au}}$, we have the long-term evolution 
\begin{equation}
\left\vert \psi \left( t\rightarrow \infty \right) \right\rangle \propto
\left( \beta _{1}\left\vert \psi _{3}\right\rangle+\beta _{2}\left\vert \psi _{4}\right\rangle \right) t^{2}.
\end{equation}

\end{document}